# Arctic teleconnection on climate and ozone pollution in the polar jet stream path of eastern US

K Shuvo Bakar[1,*], Sourish Das[2], Sudeep Shukla[3,4], Anirban Chakraborti[5,**]

[1]The University of Sydney, Camperdown, NSW 2006, Australia
[2]Chennai Mathematical Institute, Siruseri, Tamil Nadu 603103, India
[3]AI 4 Water LTD, Orpington, BR6 9QX United Kingdom
[4]Institute for Interdisciplinary Research, New Delhi 110074, India
[5]School of Computational and Integrative Sciences, Jawaharlal Nehru University, New Delhi 110067, India
[*]Email: shuvo.bakar@sydney.edu.au
[**]Email: anirban@jnu.ac.in

## Abstract

Arctic sea-ice loss is a defining feature of climate change and offers insight into its impact on mid-latitude air quality. Here, we investigate how variability in Arctic sea-ice extent (ASI) affects ground-level ozone ($O_3$) across eastern US states through physically and chemically mediated atmospheric pathways. Using observations and causal-inference methods grounded in atmospheric dynamics, we show that ASI drives wintertime ozone variability primarily via indirect meteorological mechanisms, including changes in humidity, temperature, and atmospheric circulation along the polar and subtropical jet streams. Inland regions exhibit the strongest sensitivity, while coastal areas are modulated by marine boundary-layer processes. Seasonal contrasts reveal that Arctic-driven dynamics suppress ozone in winter but can enhance accumulation under certain summer conditions. These findings highlight the importance of Arctic-midlatitude teleconnections in shaping regional air quality and highlight the need to integrate large-scale climate processes into ozone management and climate adaptation strategies.

## 1. Introduction

Arctic sea-ice extent has experienced a sustained decline over recent decades (Cai et al., 2021; Ding et al., 2017; Das et al., 2018). Climate projections indicate that even under a 1.5°C warming pathway, there remains approximately a 30% likelihood of ice-free summer conditions by the end of the century (Jahn, 2018; Screen, 2018; Sigmond et al., 2018). Variability in Arctic sea ice increasingly influences hemispheric-scale weather and climate through physically mediated teleconnections (Borzenkova et al., 2023; Cohen et al., 2018; Huang et al., 2017; Kumar et al., 2021; Vihma, 2014; Yadav et al., 2024), making Arctic sea ice a dynamically active component of the global climate system. Concurrently, global air pollution levels continue to rise in a warming climate (An et al., 2023; Ioannidis et al., 2023; Unger et al., 2006), with significant consequences for human health (Nuvolone et al., 2018) and food security (Tai et al., 2014). The coupling between large-scale climate dynamics and atmospheric composition (Tai et al., 2012; Unger et al., 2006) remains a central challenge in Earth system science, and recent work has linked Arctic sea-ice anomalies to severe haze episodes in East Asia through physically coherent circulation pathways (Yi et al., 2019; Zou et al., 2017).

Motivated by these findings, this study investigates whether teleconnection-driven process-based causality associated with Arctic sea-ice variability can influence mid-latitude air pollution, focusing on an 11-state region in the eastern United States (US). This region lies beneath the climatological influence of the polar jet stream, whose variability has been implicated in recent episodes of extreme weather and climate anomalies (Trouet et al., 2018; Wu et al., 2023). The conceptual framework guiding this work is grounded in dynamical chemical coupling, wherein large-scale atmospheric modes such as the Arctic Oscillation (AO) and North Atlantic Oscillation (NAO) modulate Arctic sea ice, which in turn alters downstream meteorological fields through changes in atmospheric thickness, thermal stratification, moisture transport, and circulation persistence.

Despite strong air-quality regulations in the United States, wintertime pollution episodes remain a recurring concern (Shah et al., 2018). We therefore examine whether winter air pollution is shaped not only by local emissions and chemistry but also by remote cryospheric forcing transmitted through large-scale atmospheric dynamics. Specifically, we focus on ground-level (tropospheric) ozone ($O_3$), a secondary pollutant with well-established impacts on respiratory and cardiovascular health (Cohen et al., 2017; Sicard, 2021). Tropospheric ozone is produced through sunlight-driven photochemical reactions involving nitrogen oxides ($NO_x$) and volatile organic compounds (VOCs) (Crutzen, 1974; Seinfeld & Pandis, 2016), and its concentration is strongly modulated by meteorological conditions such as temperature, boundary-layer structure, and relative humidity (Gorai et al., 2019; Sahu and Bakar, 2012). Within a process-based causal framework, these meteorological variables function as physical and chemical mediators linking Arctic sea-ice variability to ozone production and accumulation, while the NAO and AO act as upstream dynamical drivers influencing both sea ice and regional weather.

Air pollutants are commonly classified as primary (directly emitted) or secondary (formed through atmospheric reactions). Ozone belongs to the latter category, and its behavior is often non-linear and counterintuitive. Although elevated VOC concentrations typically promote ozone formation, wintertime ozone increases have been documented even amid declining VOC emissions (Shah et al., 2018). Similar ozone enhancements were observed during periods of reduced primary emissions associated with COVID-19 lockdowns (Anbari et al., 2022; Sicard et al., 2020). These patterns indicate that ozone levels are highly sensitive to background atmospheric composition and meteorological controls, rather than emissions alone. This complexity is further amplified by natural ozone sources, including methane-driven background ozone, wildfire emissions, and episodic stratosphere-troposphere exchange (Thompson, 2019), complicating attribution to individual drivers.

Considering these considerations, we hypothesize that the noticeable loss of Arctic sea ice (ASI) exerts a physically mediated causal influence on ground-level ozone ($O_3$) through its modification of inland meteorological conditions, particularly during winter. To evaluate this hypothesis, we employ causal-inference approaches informed by atmospheric dynamics rather than purely statistical associations. Figure 1 illustrates the proposed teleconnection-driven dynamical-chemical pathway examined in this study. We hypothesize that variability in the NAO and AO primarily governs changes in Arctic sea ice (Wang et al., 2025), which subsequently alters regional meteorological conditions including temperature, humidity, and atmospheric stability across the eastern United States (Chartrand & Pausata, 2020), hence influencing ozone formation and accumulation. In addition to this dominant pathway, direct influences of the NAO/AO on local weather, as well as potential direct effects of sea ice on ozone via circulation-driven transport and background composition, are also considered, as shown in Figure 1. This framework enables an assessment of how Arctic sea-ice loss, through its influence on the polar jet stream and its interaction with subtropical flow over the eastern United States, contributes to wintertime ozone variability while disentangling direct and indirect physical mechanisms.

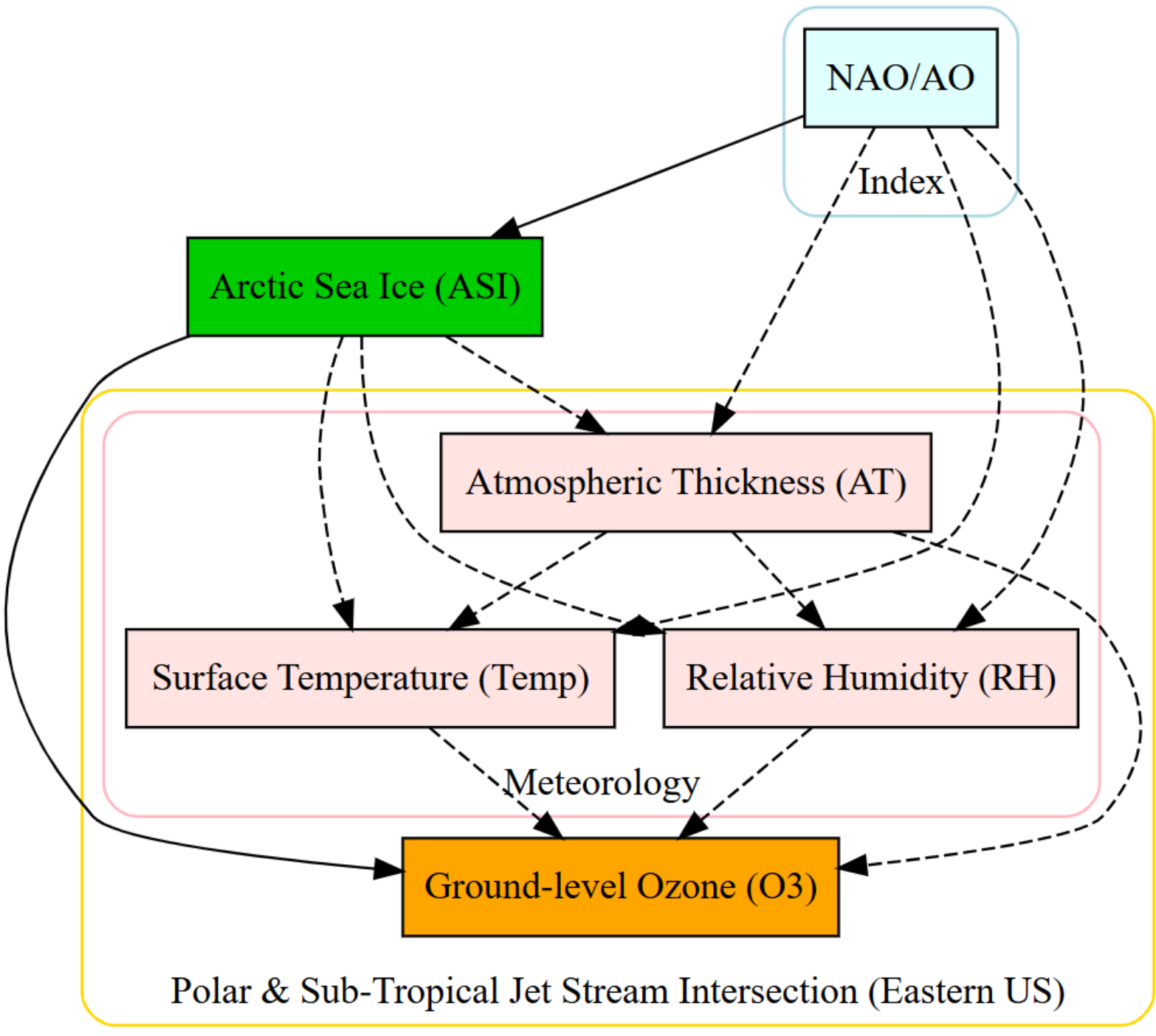

**Figure 1:** A hypothesize causal teleconnection pathway between Arctic sea ice extent and ground level ozone concentrations.

## 2. Materials and methods

### 2.1 Data source

Our study investigates the seasonal teleconnection between Arctic sea-ice extent and ground-level ozone ($O_3$) across 11 eastern U.S. states (see Figure 2) using freely available datasets covering the last three decades. Ground-level ozone data from 1989 to 2017 were collected at 154 monitoring locations in the eastern U.S., where the polar and subtropical jet streams intersect. To analyze long-term trends, we calculate monthly averages of the ozone based on the daily maximum 8-hour average concentration (in parts per billion, ppb), a widely used metric for ozone that helps assess exceedances and compliance with air-quality standards (Chameides et al., 1997; Sahu & Bakar, 2012; Sahu et al., 2007). For meteorological conditions, we consider atmospheric thickness (AT), surface temperature (Temp), and relative humidity (RH), all extracted from NOAA's NCEP reanalysis over latitudes 37.5° to 45° and longitudes -85.5° to -110° for the study period. Atmospheric thickness is calculated as the difference between geopotential heights at 1000 mb and 500 mb, while temperature (in degree C) and relative humidity (in %) are obtained from the same grid. These variables allow us to capture regional weather conditions that may mediate the influence of Arctic sea-ice variability on ozone. Monthly Arctic sea-ice extent data for 1989-2017 were downloaded from the National Snow and Ice Data Center (NSIDC), University of Colorado. Due to the limited availability of ground-level ozone measurements, the analysis is restricted to this period.

## 2.2 Methods

To examine seasonal teleconnections between Arctic sea-ice extent and ground-level ozone across 11 eastern US states (Figures 1 & 2), we apply a combination of causal-inference and statistical approaches, explicitly accounting for the chemical, meteorological, and climate processes that mediate these Arctic-midlatitude interactions. Additive model smoothing (Wood, 2008) was used to capture non-linear trends in the Arctic sea ice, meteorological and $O_3$ variables over the study period. To detect potential discontinuity years in the trends we implement a quasi-experimental framework (Hahn et al., 2001; Stock & Watson, 2015), where the assignment is based on cutoff years. From a causal analysis perspective, Figure 1 illustrates the hypothesized pathways. The NAO/AO acts as the primary upstream driver, influencing Arctic sea ice (ASI) and, to a lesser extent, regional meteorology (atmospheric thickness, surface temperature, and relative humidity). These meteorological variables serve as mediators, transmitting the influence of ASI to $O_3$. In addition to the mediated pathway, there are direct effects from NAO/AO to meteorology and from ASI to ozone. The DAG allows separation of total effects (including both mediated and direct paths) from the direct effect of ASI, while NAO/AO is treated as a confounder that must be accounted for to obtain unbiased causal estimates. Let $O_3(\mathrm{ASI}, \mathrm{AT}, \mathrm{Temp}, \mathrm{RH}, \mathrm{NAO}, \mathrm{AO})$ denote the potential ground-level ozone that would be observed if we set the variables $(\mathrm{ASI}, \mathrm{AT}, \mathrm{Temp}, \mathrm{RH}, \mathrm{NAO}, \mathrm{AO})$ to specific values. The total effect considers all causal paths from ASI to $O_3$, including indirect effects through meteorology $(\mathrm{AT}, \mathrm{Temp}, \mathrm{RH})$:

$$\text{Total Effect} = \mathrm{E}[O_3(\mathrm{ASI} = \mathrm{a}, \mathrm{NAO}, \mathrm{AO}) - O_3(\mathrm{ASI} = \mathrm{a}', \mathrm{NAO}, \mathrm{AO})]$$

Here, the total effect measures the expected change in ozone if ASI were changed from $\mathrm{a}'$ to $\mathrm{a}$, accounting for all causal pathways. To estimate this effect, we need to adjust for confounding by NAO and AO, since they affect both ASI and downstream variables:

$$\mathrm{E}[O_3 \mid \mathrm{do}(\mathrm{ASI} = \mathrm{a})] = \mathrm{E}[O_3 \mid \mathrm{ASI} = \mathrm{a}, \mathrm{NAO}, \mathrm{AO}]$$

The direct effect isolates the causal influence of ASI on $O_3$ not mediated by meteorological variables $(\mathrm{AT}, \mathrm{Temp}, \mathrm{RH})$:

$$\text{Direct Effect} = \mathrm{E}[O_3(\mathrm{ASI} = \mathrm{a}, \mathrm{AT}, \mathrm{Temp}, \mathrm{RH}, \mathrm{NAO}, \mathrm{AO}) - O_3(\mathrm{ASI} = \mathrm{a}', \mathrm{AT}, \mathrm{Temp}, \mathrm{RH}, \mathrm{NAO}, \mathrm{AO})]$$

Here, $\mathrm{AT}, \mathrm{Temp}, \mathrm{RH}$ are held fixed to block indirect pathways through meteorology, while adjusting for NAO and AO to block backdoor confounding.

To extend the causal analysis spatially and temporally, we implement a Bayesian hierarchical spatio-temporal model (Bakar & Sahu, 2015; Sahu, 2022) over spatial clusters identified via model-based clustering (Fraley & Raftery, 2002). This approach enables estimation of the seasonal and spatially varying incremental effects of ASI on ozone while accounting for local meteorological mediators and confounders. By incorporating both spatial correlation among neighboring locations and temporal dependence across months and years, the model allows us to examine how the strength of the ASI $\rightarrow$ meteorology $(\mathrm{AT}, \mathrm{Temp}, \mathrm{RH}) \rightarrow O_3$ pathway may vary across the eastern US and across different seasons. In other words, the spatio-temporal model provides a framework to quantify how the indirect effects of Arctic sea ice on ozone via meteorology differ by location and season, while also estimating direct effects and controlling for NAO/AO as a confounder. Detailed descriptions of all statistical methods are provided in the cited references, and the R software (R Core Team, 2025) code for reproducing our results is included in the Supplementary Material.

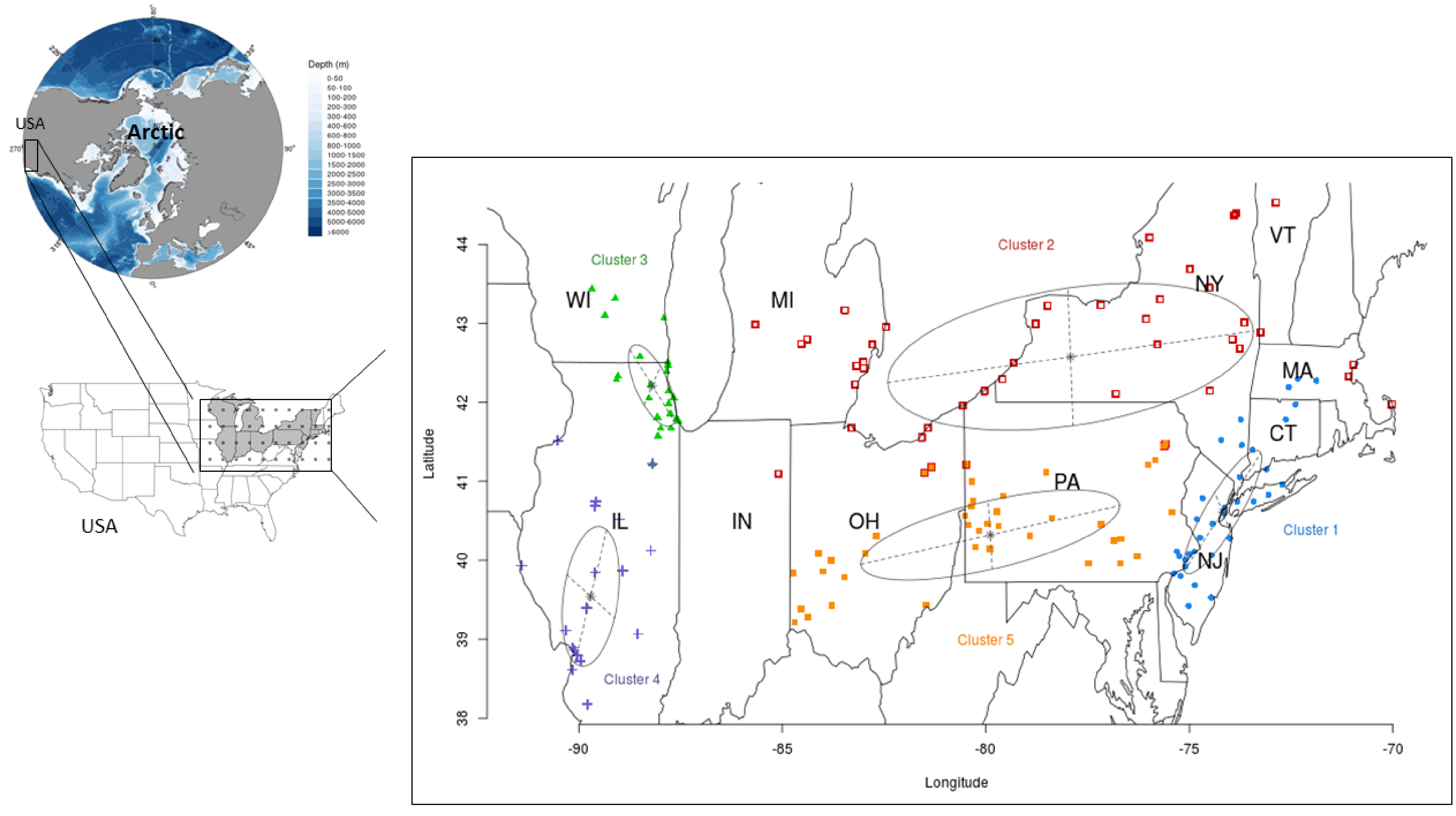

**Figure 2:** Map of the eastern United States with 154 ground level ozone monitoring stations. Five model-based spatial clusters are superimposed with different colour scheme. Grid locations for the surface air temperature and atmospheric thickness are given in US map (see bottom left panel) and Arctic Sea and US location is given in the top left panel.

## 3. Results

Arctic sea-ice extent has declined steadily in response to continued climate warming. Over the study period, sea ice decreased at an average rate of about 172 $km^2$ per day. This long-term decline is well documented in previous studies (Jahn, 2018; Screen, 2018; Sigmond et al., 2018). Seasonal analysis shows that the largest mean annual loss in sea-ice extent occurs in autumn (September-November), with a reduction of 88,177 $km^2$. The smallest loss occurs in spring (March-May), with a reduction of 34,416 $km^2$. Detailed seasonal values are provided in S1. To assess changes in Arctic sea ice from 1989 to 2017, we apply a non-linear trend analysis. The results show a strong overall decline in sea-ice extent. A noticeable downward shift appears during the early 2000s compared to earlier decades (S2). This shift may be linked to stronger seasonal climate extremes during that period. Whereas, ground-level ozone shows an upward shift over the same time frame (S2). To further examine this behavior, we perform a discontinuity analysis with a placebo sensitivity test for the period 2000-2007. We select 2007 as the endpoint because variability in the non-linear trend decreases afterward. The results reveal a strong and statistically significant discontinuity in Arctic sea-ice extent in 2001 (S3).

The influence of Arctic sea-ice extent on ground-level ozone across the eastern US shows clear spatial differences. These patterns are strongest in regions aligned with the intersection of the polar and subtropical jet streams. Using spatial clustering and hierarchical modeling, we identify distinct regions where Arctic sea-ice variability has a statistically significant effect on surface ozone (Figure 2). Ozone seasonality across monitoring sites shows relatively stable concentrations between 38° and 42° latitude during 1989-2017 (S4). Whereas sites north of 42° latitude show much greater variability, driven mainly by increases in wintertime ozone. Longitudinal differences in ozone are smaller and weaker than latitudinal differences (S4). We identify five spatial clusters across the eastern US, each located along the jet stream corridor. In all clusters, continued Arctic sea-ice loss is associated with a negative and statistically significant effect on ground-level ozone. The strength of this effect varies by region. The strongest response occurs west of Lake Michigan (cluster 3). The weakest response is found east of Lake Michigan, including northern New York and Vermont (cluster 2) (S5). When spatio-temporal correlation is included, the results become stronger and more consistent. In spring, a similar negative relationship is observed in most regions. However, the effect is not statistically significant in areas covering Michigan, the Lake Erie shoreline, northern New York, and Vermont (cluster 2). In autumn, Arctic sea-ice extent has a statistically significant negative effect on ozone in New Jersey, Long Island (New York), Connecticut, and western Massachusetts (cluster 1), as well as in southern Pennsylvania and Ohio (cluster 5). In summer, statistically significant negative effects persist in Wisconsin and Illinois (clusters 3 and 4).

Table 1 summarizes the causal pathways through which Arctic sea-ice extent (ASI) influences ground-level ozone ($O_3$) across the eastern US, distinguishing between indirect (mediated) effects and direct dynamical effects and resolving these relationships by season and region. Negative values indicate that reductions in Arctic sea-ice extent are associated with higher ground-level ozone concentrations, whereas positive values indicate that reductions in Arctic sea-ice are associated with lower ozone concentrations. This interpretation reflects how Arctic-driven meteorological and chemical processes modulate ozone accumulation near the surface. The indirect effects, represented by the average causal mediation effects (ACME), show that meteorological and chemical pathways dominate the Arctic sea-ice and ozone relationship. Among these pathways, the relative humidity ($ASI \rightarrow RH \rightarrow O_3$) mediation effect is consistently negative and has the largest magnitude across nearly all seasons and spatial clusters. This indicates that Arctic sea-ice loss tends to increase ground-level ozone through humidity-related processes. Physically, reduced sea ice alters moisture transport and synoptic circulation in ways that can decrease cloud cover, reduce wet removal, and limit heterogeneous chemical loss of ozone and its precursors, hence favoring ozone accumulation. These effects are strongest in the northeastern US and the Ohio-Pennsylvania region, highlighting the importance of humidity-driven chemical processes in regulating surface ozone.

The temperature-mediated pathway ($\mathrm{ASI} \rightarrow \mathrm{Temp} \rightarrow O_3$) is weaker than the humidity pathway and exhibits seasonal dependence. In summer, the ACME via temperature is often negative, indicating that Arctic sea-ice loss promotes higher ozone concentrations through temperature-related enhancements in photochemical production. This is consistent with warmer conditions favoring faster reaction rates and increased ozone formation during photochemically active seasons. In winter, temperature-mediated effects are small and sometimes positive, reflecting the limited role of temperature-driven photochemistry under low solar radiation and cold conditions. Overall, temperature acts as a secondary mediator compared to relative humidity. The atmospheric thickness pathway ($\mathrm{ASI} \rightarrow \mathrm{AT} \rightarrow O_3$) reflects changes in large-scale thermal structure and atmospheric stability. This pathway shows mixed seasonal behavior, with negative values dominating in winter and positive values more common in summer and autumn. Negative winter values indicate that Arctic sea-ice loss is associated with higher ozone through changes in atmospheric structure that favor stagnation or reduced vertical mixing, allowing ozone to accumulate near the surface. These effects are more pronounced in inland clusters, where large-scale circulation changes exert stronger control over surface air quality than local boundary-layer processes. When all mediated pathways are combined, the total ACME is negative and statistically significant across nearly all regions and seasons. This result indicates that, overall, reductions in Arctic sea-ice are associated with higher ground-level ozone concentrations through indirect meteorological and chemical pathways. The strongest mediated effects occur in winter and in inland Midwestern regions, highlighting the sensitivity of continental interiors to Arctic-driven changes in circulation, moisture transport, and atmospheric stability.

In addition to mediated effects, Table 1 reports the average direct effect (ADE) of Arctic sea-ice on ozone, representing influences that are not explained by temperature, humidity, or atmospheric thickness. These direct effects show strong seasonal contrasts. During winter and spring, the ADE is strongly negative across all clusters, indicating that Arctic sea-ice loss directly increases ground-level ozone concentrations. This finding implies that higher Arctic sea-ice is associated with lower wintertime ozone, while reduced sea ice favors conditions such as weaker ventilation, altered jet-stream positioning, and enhanced persistence of stagnant air masses that promote ozone accumulation. These direct effects are strongest in inland regions aligned with the polar jet stream. During summer, the direct Arctic sea-ice and ozone relationship becomes more regionally heterogeneous. In the northeastern coastal region, the ADE is positive, suggesting that Arctic sea-ice loss may reduce ozone through enhanced ventilation or altered transport pathways. This response is largely absent in inland clusters, where summer ozone remains primarily controlled by local emissions and photochemical production. Together, these results demonstrate that ground-level ozone responds to Arctic sea-ice variability through a combination of chemical, meteorological, and climate-driven mechanisms, with indirect humidity-related pathways and direct circulation effects playing particularly important roles during winter.

**Table 1:** Seasonal and Regional Causal Effects of Arctic Sea-Ice Extent on Ground-Level Ozone via Temperature, Atmospheric Thickness, and Relative Humidity in the eastern US polar jet stream path.

| **Effect** | **Cluster 1** | **Cluster 2** | **Cluster 3** | **Cluster 4** | **Cluster 5** | **All** |
|---|---|---|---|---|---|---|
| **Average Causal Mediation Effect (ACME) via Surface Temperature** | | | | | | |
| Summer | -1.298* | -0.800* | -1.734* | -1.789* | -1.714* | -1.149* |
| Autumn | -0.653* | -0.781* | -0.478* | -0.540* | -1.374* | -0.717* |
| Winter | -0.226* | 0.245* | 0.195* | 0.336* | 0.271* | -0.041 |
| Spring | 0.245* | -0.048 | 0.114 | -0.327* | -0.183* | -0.030* |
| All | -0.084* | 0.028 | 0.057 | -0.296* | -0.590* | -0.160* |
| **Average Causal Mediation Effect (ACME) via Atmospheric Thickness** | | | | | | |
| Summer | 0.119 | 0.024 | 0.502* | 0.618* | 0.369* | 0.199* |
| Autumn | 0.187* | 0.357* | 0.199* | 0.199* | 0.553* | 0.295* |
| Winter | -0.161* | 0.093* | 0.005 | -0.032 | -0.023* | -0.058* |
| Spring | 0.098* | -0.059* | -0.246* | -0.013 | 0.188* | 0.017 |
| All | 0.093* | 0.001 | -0.168* | 0.136* | 0.253* | 0.067* |
| **Average Causal Mediation Effect (ACME) via Relative Humidity** | | | | | | |
| Summer | -0.283* | -0.047* | -0.134* | -0.375* | -0.289* | -1.149* |
| Autumn | -1.988* | -0.089* | 0.270* | -0.029 | -0.083* | -0.048* |
| Winter | -0.568* | -0.625* | -0.306* | -0.624* | -1.085* | -0.001 |
| Spring | -0.619* | -0.427* | -1.125* | -0.987* | -0.875* | -0.704* |
| All | -0.287* | -0.230* | -0.042* | -0.346* | -0.388* | -0.237* |
| **Total Causal Mediation Effect** | | | | | | |
| Summer | -1.462* | -0.823* | -1.366* | -1.546* | -1.633* | -1.131* |
| Autumn | -0.665* | -0.513* | -0.009* | -0.368* | -0.904* | -0.470* |
| Winter | -0.955* | -0.288* | -0.106* | -0.320* | -0.836* | -0.099* |
| Spring | -0.277* | -0.535* | -1.257* | -1.327* | -0.869* | -0.717* |
| All | -0.278* | -0.201* | -0.154* | -0.507* | -0.725* | - 0.330* |
| **Average Direct Effect (ASI ‣ $O_3$)** | | | | | | |
| Summer | 2.169* | 1.105* | -0.453* | 0.103 | 1.531* | 0.925* |
| Autumn | -0.694* | -0.463* | -1.748* | 0.976* | -1.191* | -1.038* |
| Winter | -3.640* | -4.118* | -4.403* | -3.756* | -6.571* | -4.977* |
| Spring | -4.118* | -1.290* | -1.742* | -1.276* | -3.532* | -2.394* |
| All | -0.950* | -0.662* | -1.951* | -1.186* | -1.172* | -1.420* |
| **Total Direct Effect (ASI ‣ $O_3$)** | | | | | | |
| Summer | 0.707* | 0.282* | -1.819* | -1.443* | -0.102* | -0.206* |
| Autumn | -1.359* | -0.976* | -1.757* | 0.608* | -2.095* | -1.508* |
| Winter | -4.595* | -4.406* | -4.509* | -4.076* | -7.407* | -5.076* |
| Spring | -4.395* | -1.825* | -2.999* | -2.603* | -4.401* | -3.111* |
| All | -1.228* | -0.863* | -2.105* | -1.693* | -1.897* | -1.516* |

## 3.1 Projection scenario in 2030

We also explore possible projection scenario trends in both Arctic sea-ice extent and ozone concentrations under the model assumption. As expected, we see that the sea-ice extent is decreasing over the years (S6). The projected Arctic sea-ice extent in 2030 for winter months (i.e., Dec-Feb) is $11.53 \times 10^6$ square KM, whereas, in 2010 winter months, it was $13.38 \times 10^6$ square KM. Additionally, we notice a sudden decremental non-linear trend after 2020, which is almost similar for all seasons. We also estimate the model-based ozone concentration trends, see Figure 3. The spatially distributed overall trend of ozone concentrations exhibits an increasing pattern in most of the parts of the eastern US polar jet stream pathway, such as western part of Pennsylvania, northern part of Indiana and Illinois. However, north-western parts of Wisconsin and north-eastern part of New York shows a decreasing projected trend, see Figure 3(a). Season-wise, we see that the ozone trends in summer months (i.e., June to August) are decreasing for almost whole eastern US, except few city areas in Illinois and Michigan, see Figure 3(d). Whereas, the winter ozone trend shows an increasing pattern throughout our study region, see Figure 3(b).

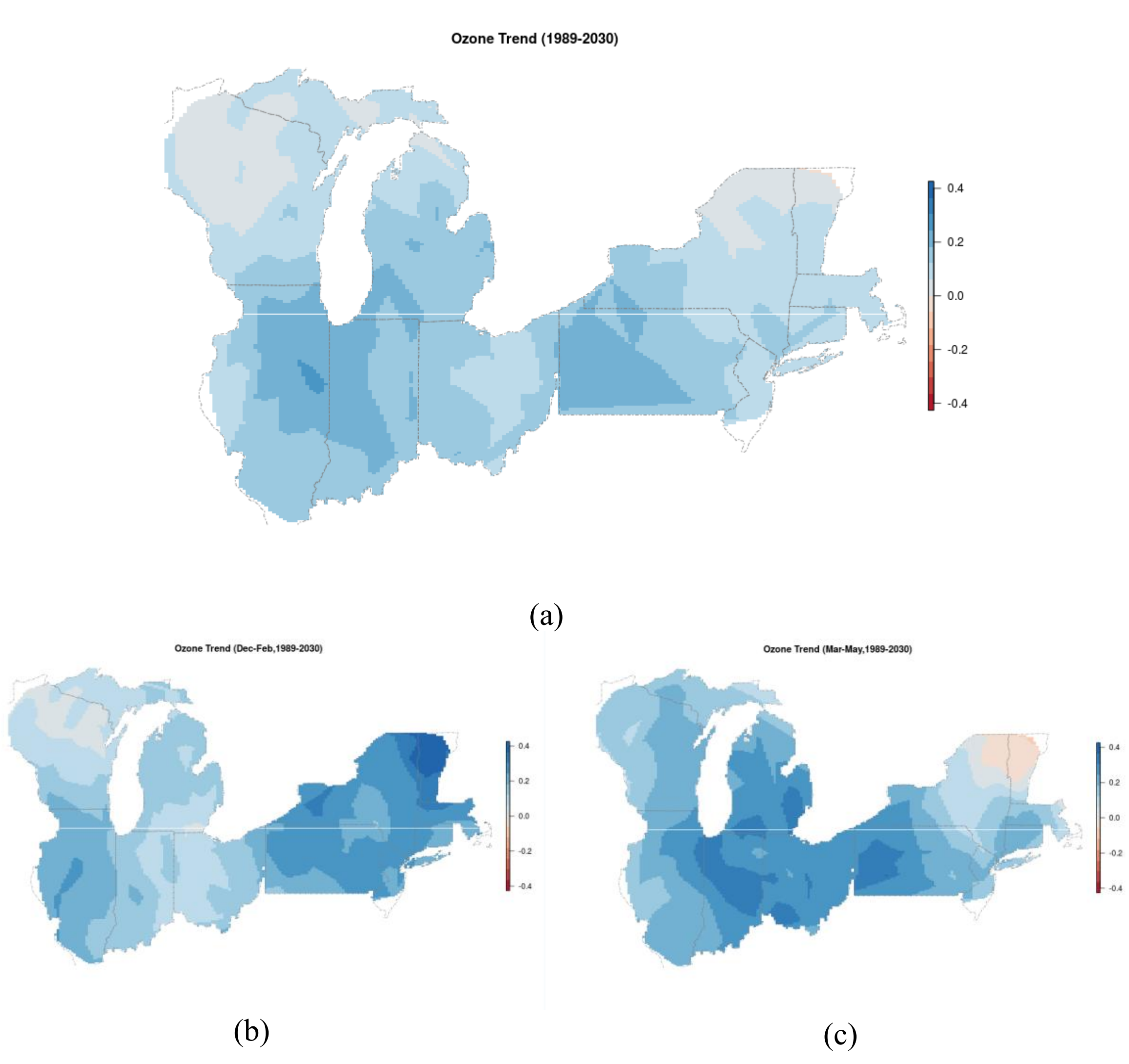

(a)

(b) (c)

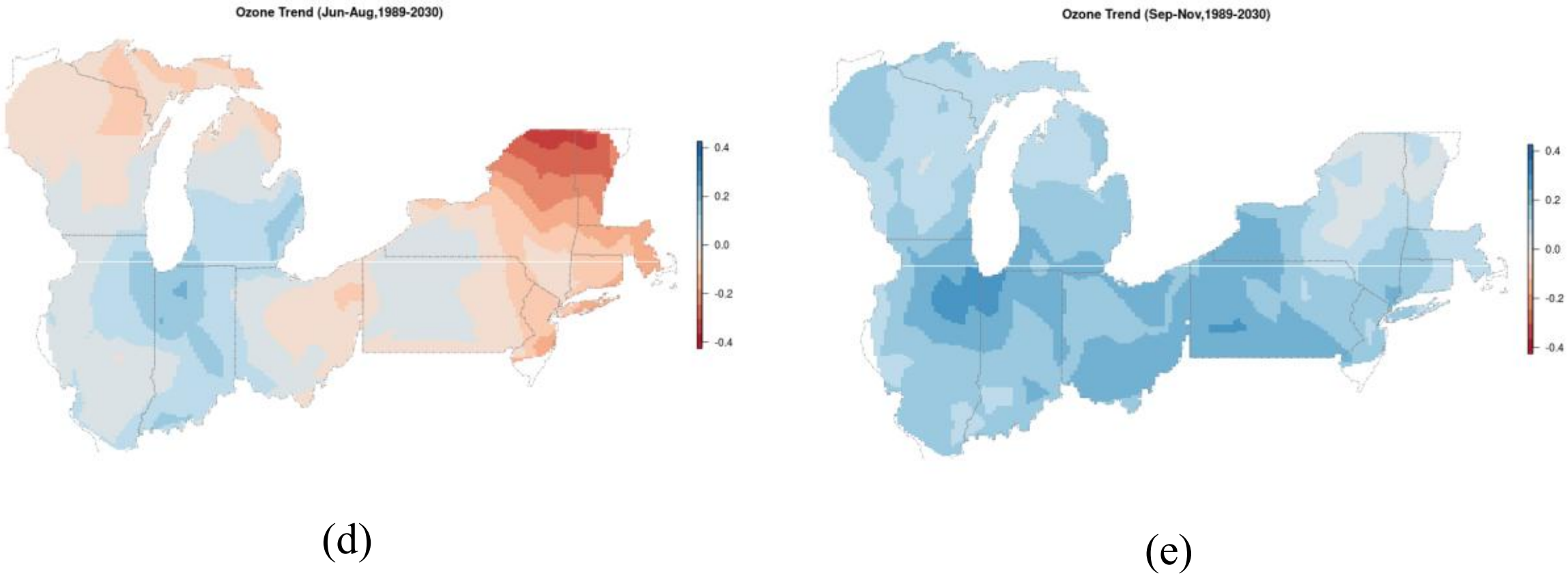

**Figure 3:** Projected trend (1989-2030) in ground level ozone concentrations. The top panel (a) represents overall ozone trend in the eastern US polar jet stream path. Middle and bottom panels represent ozone trends by seasons: (b) Winter (Dec-Feb), (c) Spring (Mar-May), (d) Summer (Jun-Aug) and (e) Autumn (Sep-Nov).

## 4. Discussion

This study shows that Arctic sea-ice loss significantly affects ground-level ozone concentrations across the eastern US through interconnected chemical, meteorological, and climate mechanisms. Arctic sea ice influences regional air quality through direct dynamical effects and indirectly by changing local temperature, humidity, and atmospheric stability, which in turn regulate ozone formation and removal. From a chemical perspective, relative humidity is the dominant pathway linking Arctic sea-ice changes to ozone. Lower Arctic sea ice alters moisture transport, which reduces cloud cover, limits wet deposition, and decreases heterogeneous chemical loss of ozone and its precursors. These processes allow ozone to accumulate near the surface, especially in winter. Temperature also mediates ozone formation, mainly during summer, when higher temperatures promote faster photochemical reactions and increased ozone production. In winter, chemical production is limited, so ozone behavior is more sensitive to transport and accumulation rather than local photochemistry. From a meteorological perspective, reductions in Arctic sea ice weaken the temperature gradient between the Arctic and mid-latitudes, which changes the position and strength of the polar jet stream. These circulation shifts influence wind patterns, vertical mixing, and ventilation, particularly over inland regions. In winter, weaker ventilation and more stagnant conditions allow ozone to build up, while in summer, local photochemistry and emissions dominate ozone variability. Atmospheric thickness also contributes by affecting vertical mixing and stability, allowing ozone to accumulate when mixing is limited. Coastal regions show different responses due to land-sea contrasts and boundary-layer effects. From a climate perspective, the long-term decline in Arctic sea ice represents a persistent shift in the Earth system that has far-reaching effects on mid-latitude air quality. The timing of sea-ice loss aligns with changes in ground-level ozone, suggesting a physical coupling between Arctic climate variability and regional ozone patterns. Regions along the polar jet stream corridor, particularly inland Midwestern areas, are especially sensitive to these climate-driven changes, highlighting the role of large-scale Arctic-midlatitude teleconnections in shaping seasonal ozone variability.

These findings have important policy and air-quality implications. Traditional ozone management strategies focus on local emissions and summer ozone episodes, but Arctic-driven changes can increase wintertime ozone in regions that have not historically experienced high pollution. Incorporating Arctic climate indicators, such as sea-ice extent, into forecasting and regulatory frameworks could improve preparedness and mitigation strategies. Reducing greenhouse gas and short-lived climate pollutant emissions may have dual benefits by limiting Arctic warming and helping control background ozone levels. Effective ozone management requires integrating chemical, meteorological, and climate drivers to anticipate and respond to evolving air-quality challenges in a warming world.

## 5. Conclusion

Arctic sea-ice loss affects ground-level ozone in the eastern US through chemical, meteorological, and climate-related processes. Reduced sea ice alters temperature, humidity, and atmospheric circulation, which can increase ozone concentrations, especially during winter in inland regions. Summer ozone is less sensitive to Arctic changes but can still be affected in coastal areas through circulation and boundary-layer effects. These findings show that Arctic climate variability plays a key role in shaping mid-latitude air quality. Incorporating these insights into air-quality management and policy can improve forecasting, guide emission controls, and help mitigate the impacts of climate change on regional ozone pollution.

**Supplementary:**

**S1:** Arctic sea-ice extent trends by season over years 1989-2017.

| Season | Intercept | Year | Ice-lost per day |
|---|---|---|---|
| **Dec-Feb** | 121.6151* | -0.0538* | 147.43 $km^2$ |
| **Mar-May** | 83.1140* | -0.0344* | 94.29 $km^2$ |
| **Jun-Aug** | 159.7561* | -0.0753* | 206.27 $km^2$ |
| **Sep-Nov** | 184.5781* | -0.0882* | 241.58 $km^2$ |
| **All** | 137.1625* | -0.0689* | 172.27 $km^2$ |

‘*’ implies statistically significant at 5% level of significance. Model estimates not adjusted by spatial correlation.

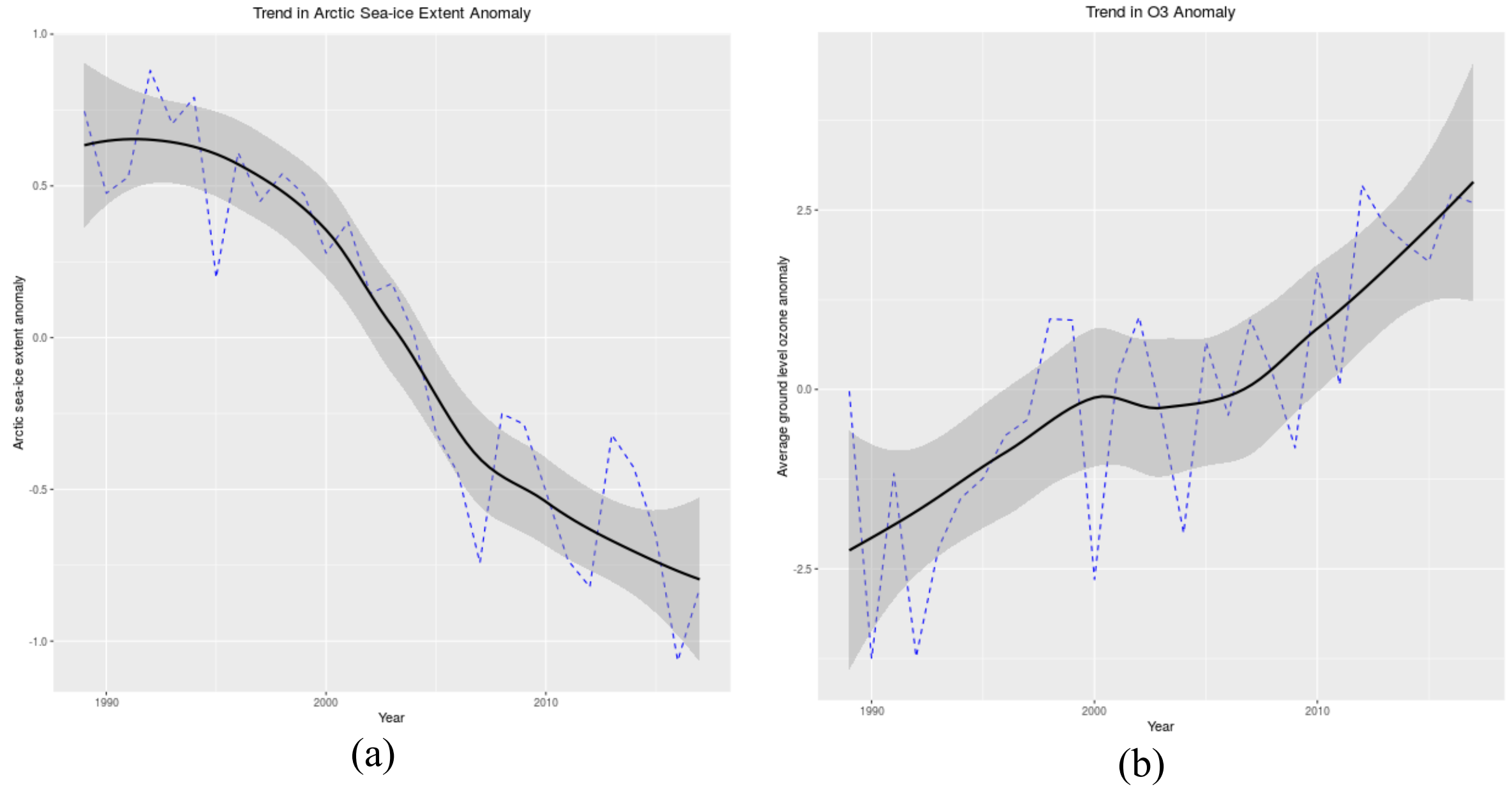

**S2:** Trends in (a) Arctic sea-ice extent and (b) ground level $O_3$ concentrations. Trends are calculated from the anomalies of the corresponding variables. The actual year-wise anomalies are also superimposed on the plots (blue dashed lines).

**S3:** Sensitivity analysis of the discontinuity effect using Regression Discontinuity Design (RDD) by years from 2000 to 2006 for Arctic sea-ice extent, air temperature, atmospheric thickness and ground level ozone concentrations (scaled observations).

| Discontinuity year | Arctic | Ozone |
|---|---|---|
| 2000 (1989-1999 vs. 2000-2017) | -0.09 | -2.06* |
| 2001 (1989-2000 vs. 2001-2017) | **-0.14*** | -0.85* |
| 2002 (1989-2001 vs. 2002-2017) | -0.28* | -1.01* |
| 2003 (1989-2002 vs. 2003-2017) | -0.33* | -1.57* |
| 2004 (1989-2003 vs. 2004-2017) | -0.43* | **-1.22*** |
| 2005 (1989-2004 vs. 2005-2017) | -0.46* | 0.14 |
| 2006 (1989-2005 vs. 2006-2017) | -0.35* | 0.08 |
| 2007 (1989-2006 vs. 2007-2017) | -0.20* | 0.65* |

'*' implies statistically significant at 5% level of significance.

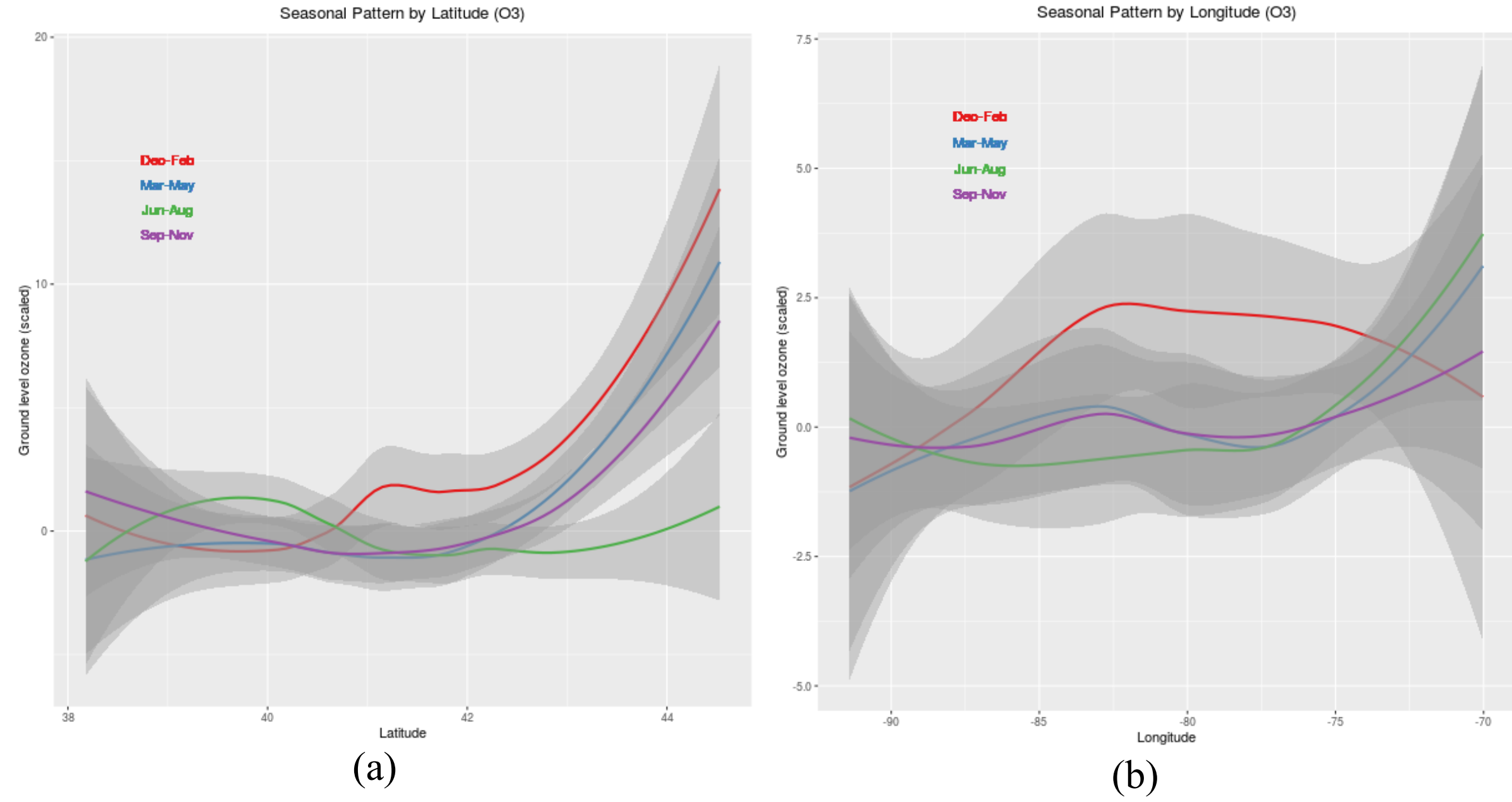

**S4:** Non-linear trends in ground level ozone concentrations in eastern US study region explored by seasons and by spatial (a) latitude and (b) longitue positions.

**S5:** Effects of Arctic sea-ice extent on ground level ozone concentrations by spatial clusters.

| Regression model parameter estimates (scaled 1989-2017 and without adjusting spatio-temporal correlation) | | | | | | |
|---|---|---|---|---|---|---|
| | | **Cluster 1** | **Cluster 2** | **Cluster 3** | **Cluster 4** | **Cluster 5** |
| **Intercept** | | 0.13* | 1.49* | -1.91* | -0.37* | -0.71* |
| **Arctic sea-ice extent effect** | | -1.24* | -0.93* | -2.43* | -1.65* | -1.92* |
| Hierarchical Bayesian model (HBM) parameter estimates (scaled 1989-2017 and with an adjustment of spatio-temporal correlation) | | | | | | |
| | | **Cluster 1** | **Cluster 2** | **Cluster 3** | **Cluster 4** | **Cluster 5** |
| **Intercept** | | | | | | |
| | Summer | 2.36* | 2.48 | -1.53* | 0.08 | -0.15 |
| Incremental effects compared to Summer | | | | | | |
| | Autumn | -1.80* | 1.17 | 0.42 | 0.61 | -1.64* |
| | Winter | -2.68* | 3.78* | -0.45 | -1.74* | -3.44* |
| | Spring | 0.38 | 1.37 | 0.11 | 0.69* | -1.02* |
| **Arctic sea-ice effect** | | | | | | |
| | Summer | 1.23* | 1.56 | -1.73* | -0.61* | -0.28 |
| Incremental effects compared to Summer | | | | | | |
| | Autumn | -2.34* | -2.06 | -0.57 | -0.62 | -1.78* |
| | Winter | -5.74* | -6.06* | -2.85* | -4.35* | -5.38* |
| | Spring | -4.36* | -2.78 | -2.59* | -2.24* | -3.88* |
| **Temporal auto-regressive parameter** | | | | | | |
| | ρ | 0.21* | 0.56* | 0.29* | 0.09* | 0.11* |
| **Spatial range (in KM)** | | | | | | |
| $-\log(0.05)/\phi$ | | 460.88* | 1361.69* | 329.21* | 491.10* | 855.92* |
| '*' implies statistically significant at 5% level of significance. | | | | | | |

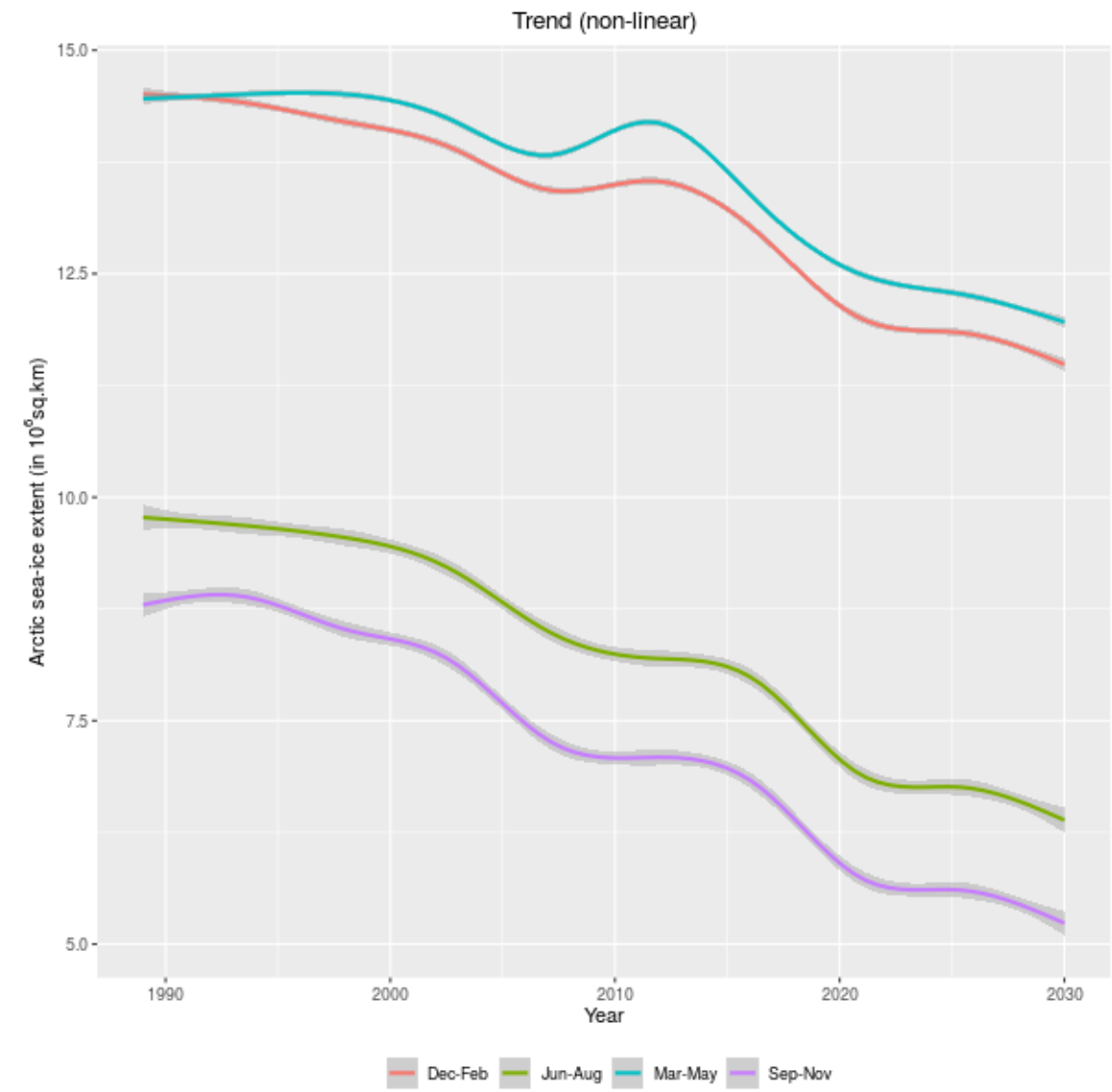

**S6:** Model based forecast of Arctic sea-ice extent till 2030, with corresponding non-linear trends by seasons.